\documentclass[preprint,aps,prc,superscriptaddress,floatfix]{revtex4-1}
\usepackage{graphicx}
\usepackage{dcolumn}
\usepackage{longtable}

\begin{document}

\title{Multi-step processes in heavy-ion induced single-nucleon transfer reactions}

\author{N. Keeley}
\email[Corresponding author: ]{nicholas.keeley@ncbj.gov.pl}
\affiliation{National Centre for Nuclear Research, ul.\ Andrzeja So\l tana 7, 05-400
Otwock, Poland}
\author{K. W. Kemper}
\affiliation{Department of Physics, Florida State University, Tallahassee, Florida 32306, USA}
\affiliation{Heavy Ion Laboratory, University of Warsaw, ul.\ Pasteura 5a, 02-093 Warsaw, Poland}
\author{K. Rusek}
\affiliation{Heavy Ion Laboratory, University of Warsaw, ul.\ Pasteura 5a, 02-093 Warsaw, Poland}

\begin{abstract}
It was first noted during the 1970s that finite-range distorted wave Born approximation (FR-DWBA) calculations 
were unable satisfactorily to describe the shape of the angular distributions of many single-proton 
(and some single-neutron) transfer reactions induced by heavy ions, with calculations shifted 
to larger angles by up to $\sim 4^\circ$ compared with the data.
These reactions exhibited a significant mismatch, either of the reaction Q value or the grazing angular momentum of the
entrance and exit channels, and it was speculated that the inclusion of multi-step transfer paths via excited state(s)
of the projectile and/or ejectile could compensate for the effect of this mismatch and yield good descriptions of 
the data by shifting the calculated peaks to smaller angles. However, to date
this has not been explicitly demonstrated for many reactions. In this work we show that inclusion of the two-step transfer
path via the 4.44-MeV $2^+$ excited state of the $^{12}$C projectile in coupled channel Born approximation calculations
enables a good description of the $^{208}$Pb($^{12}$C,$^{11}$B)$^{209}$Bi single-proton stripping data at four
incident energies which could not be described by the FR-DWBA. We also show that inclusion of a similar reaction 
path for the $^{208}$Pb($^{12}$C,$^{13}$C)$^{207}$Pb
single-neutron pickup reaction has a relatively minor influence, slightly improving the already good description obtained
with the FR-DWBA.
\end{abstract}

\maketitle

\section{Introduction}\label{intro}
When reactions induced by heavy ions were first extensively studied in the 1970s, it was found that while
the single-neutron transfer data could usually be well described by finite-range distorted wave Born approximation
(FR-DWBA) calculations, in many systems the single-proton transfer data showed significant angular shifts
compared to the calculations. A good example of this phenomenon was seen in the $^{12}$C + $^{208}$Pb 
system where the $^{208}$Pb($^{12}$C,$^{13}$C)$^{207}$Pb single-neutron pickup data were well described
by FR-DWBA calculations, whereas similar calculations for the $^{208}$Pb($^{12}$C,$^{11}$B)$^{209}$Bi 
single-proton stripping showed a progressively greater shift to larger angles (typically a few degrees)
compared to the data as successively higher-lying levels in $^{209}$Bi were populated \cite{Tot76}.

Data for three near-barrier energies, 77.4, 97.9 and 116.4 MeV, are presented in Ref.\ \cite{Tot76}.
For the proton-stripping reaction the discrepancy between FR-DWBA calculations and data increases as 
the $Q$ value becomes more negative, i.e.\ as more highly excited states of the $^{209}$Bi residual
are populated, or the bombarding energy is reduced, while for the neutron pickup reaction the data
are well described by FR-DWBA calculations for all states of the $^{207}$Pb residual at all bombarding 
energies. Data for the same reactions at a bombarding energy of 101 MeV exhibit similar behavior \cite{Oer84}. 

The difference in behavior of the proton stripping and neutron pickup reactions in the $^{12}$C + $^{208}$Pb
system may be understood by means of the concept of ``matching.'' For systems involving heavy ions as
projectiles at energies close to the Coulomb barrier the Sommerfeld parameter is large and a semiclassical 
picture based on Rutherford trajectories should hold. Within this view a transfer reaction will only
take place with appreciable probability if the trajectories before and after the transfer are continuous.
Imposing this condition leads to matching criteria for the reaction Q value, see e.g.\ Buttle and
Goldfarb \cite{But71}. A more detailed understanding of matching requirements may be obtained by considering
continuity of linear and angular momenta in the initial and final trajectories and in a seminal article
Brink \cite{Bri72} gave matching conditions for the angular momentum transfer and Q value based on these
criteria which should be satisfied if the transfer probability is to be large.

These matching conditions lead to the concept of the ``Q window'' for heavy ion reactions, whereby those
levels with Q values within a few MeV of the optimum value are seen to be preferentially populated. 
Transfer to a given level must also satisfy the angular momentum matching requirements in that the
difference between the grazing angular momenta of the incoming and outgoing trajectories should closely match the
allowed angular momentum transfers. In Ref.\ \cite{Tot76} it was noted that the discrepancy between the
FR-DWBA calculations and the data for the $^{208}$Pb($^{12}$C,$^{11}$B)$^{209}$Bi reactions was
correlated with the degree of mismatch between the grazing angular momenta in the entrance and exit channels;
in other words, the angular momentum mismatch tended to be larger for the population of levels in $^{209}$Bi at higher 
excitation energies and lower incident $^{12}$C energies, exactly where the discrepancy between FR-DWBA calculations
and data was largest. The $^{208}$Pb($^{12}$C,$^{13}$C)$^{207}$Pb reaction satisfies the matching requirements rather better,
hence the FR-DWBA was able to provide a good description of these data.

It was speculated in both Refs.\ \cite{Tot76} and \cite{Oer84} that the shift in angle of the FR-DWBA
calculations compared to the $^{208}$Pb($^{12}$C,$^{11}$B)$^{209}$Bi 
data could be due to the effects of couplings to inelastic channels, particularly
of the lighter reaction partner. However, this has not to date been explicitly demonstrated. 
Given the renewed interest in reactions induced by light heavy ions in connection with the availability of radioactive
beams it seems timely to revisit the existing data for stable systems where these problems occur using more
sophisticated reaction models. In this work we show
that it is possible to obtain a good description of the $^{208}$Pb($^{12}$C,$^{11}$B)$^{209}$Bi data 
at all four bombarding energies without the need for {\em ad hoc} adjustments of the optical potentials by 
including the two-step transfer path via the 4.44-MeV $2^+_1$ state of $^{12}$C in coupled channel Born
approximation (CCBA) calculations. We also demonstrate that the inclusion of this reaction path in similar
calculations for the $^{208}$Pb($^{12}$C,$^{13}$C)$^{207}$Pb reaction has a relatively small effect, slightly
improving the already good description of these data by FR-DWBA calculations.

\section{Calculations}\label{calcs}
\subsection{The $^{208}$Pb($^{12}$C,$^{11}$B)$^{209}$Bi proton stripping reaction}\label{calcs:p}
The data sets of Refs.\ \cite{Tot76} and \cite{Oer84} were reanalyzed.
A series of CCBA calculations of the $^{208}$Pb($^{12}$C,$^{11}$B)$^{209}$Bi proton stripping reaction including 
the two-step transfer path via the 4.44-MeV $2^+$ excited state of $^{12}$C were performed with the code {\sc fresco} 
\cite{Tho88}, employing the post form of the DWBA formalism and incorporating the full complex remnant term. 
Inputs were similar to the FR-DWBA calculations of Ref.\ \cite{Tot76}. The most extensive 
data set is that at a $^{12}$C bombarding energy of 97.9 MeV so this was used as a test case. In the
entrance partition, coupling to the 4.44-MeV $2^+_1$ state of $^{12}$C was included using standard rotational
model form factors for an oblate quadrupole deformation. The $B(E2; 0^+ \rightarrow 2^+)$ was taken from Ref.\
\cite{Ram01} and the nuclear deformation length, $\delta_2 = -1.40$ fm, from Ref.\ \cite{Coo86}. The optical
potential was of Woods-Saxon form and the parameters were adjusted so that the coupled channel calculation 
gave the same elastic scattering angular distribution as an optical model calculation using the $^{12}$C + $^{208}$Pb 
parameters listed in Table I of Ref.\ \cite{Tot76}. The resulting values are given in Table \ref{tab:omp}. 
\begin{table}
\begin{center}
\begin{tabular}{|c|c|c|c|c|c|c|}
\hline
$^{12}$C energy (MeV) & $V$ (MeV) & $r_0$ (fm) & $a_0$ (fm) & $W$ (MeV) & $r_W$ (fm) & $a_W$ (fm) \\ \hline
77.4 & 52.2 & 1.256 & 0.56 & 5.02 & 1.256 & 0.792 \\
97.9 & 49.9 & 1.256 & 0.56 & 11.7 & 1.256 & 0.406 \\
101.0 & 51.1 & 1.256 & 0.56 & 13.3 & 1.256 & 0.415 \\
116.4 & 42.0 & 1.256 & 0.56 & 20.0 & 1.256 & 0.406 \\
\hline
\end{tabular}
\caption{Entrance channel $^{12}$C + $^{208}$Pb Woods-Saxon potential parameters used in the CCBA calculations.
The heavy-ion radius convention, $R_i = r_i \times (\mathrm{A}_p^{1/3} + \mathrm{A}_t^{1/3})$ fm was used and 
$r_C = 1.30$ fm at all energies. The larger imaginary diffuseness at 77.4 MeV is required to fit the elastic
scattering data of Ref.\ \cite{Rud01}.}
\label{tab:omp}
\end{center}
\end{table}
The five-parameter $^{11}$B + $^{209}$Bi Woods-Saxon potential from Table I of Ref.\ \cite{Tot76} was employed in the exit
channel. This was obtained by fitting $^{11}$B + $^{209}$Bi elastic scattering data at an appropriate energy (74.6 MeV).
The $^{11}\mathrm{B} + p$ and $^{208}\mathrm{Pb} + p$ binding potentials were taken from Ref.\ \cite{Tot76}.

Two sets of spectroscopic amplitudes for the $\left<^{12}\mathrm{C}(0^+) \mid \protect{^{11}\mathrm{B}} + p \right>$ and
$\left<^{12}\mathrm{C}(2^+) \mid \protect{^{11}\mathrm{B}} + p \right>$ overlaps were tested, those of Cohen and Kurath
\cite{Coh67} and those of Rudchik {\em et al.\/} \cite{Rud01a}. The values are given in Table \ref{tab2}. The spectroscopic
amplitudes for the $\left<^{209}\mathrm{Bi} \mid \protect{^{208}\mathrm{Pb}} + p \right>$
overlaps were adjusted to give the best description of the stripping cross sections. The small $1p_{3/2}$ component of
the $\left<^{12}\mathrm{C}(2^+) \mid \protect{^{11}\mathrm{B}} + p \right>$ overlap predicted by Cohen and Kurath \cite{Coh67}
was omitted from the calculations since tests found that it had no discernible effect on the results. Since Cohen and Kurath
\cite{Coh67} give values for the spectroscopic factors, which are positive definite, and we require spectroscopic amplitudes, which
may be positive or negative, for our CCBA calculations, both positive and negative relative signs of the spectroscopic amplitudes
for the $\left<^{12}\mathrm{C}(0^+) \mid \protect{^{11}\mathrm{B}} + p \right>$ and $\left<^{12}\mathrm{C}(2^+) \mid
\protect{^{11}\mathrm{B}} + p \right>$ overlaps were tested. However, there was a clear preference for a negative relative
sign, since a positive sign was found to shift the peaks in the calculated stripping angular distributions to larger
angles, worsening the agreement with the data; a negative sign shifted the calculated peaks to smaller angles. Rudchik {\em et
al.\/} \cite{Rud01a} give the spectroscopic amplitudes so these were used directly, although since the phase convention adopted 
in Table 2 of Ref.\ \cite{Rud01a} is different from that of {\sc fresco} the signs first had to be converted. The signs
in Table \ref{tab2} follow the {\sc fresco} convention, and the conversion factor is given in the caption.
\begin{table}
\begin{center}
\begin{tabular}{|c|c|c|c|c|}
\hline
A & C & $x$ & $n\ell_j$ & $\mathrm{S}_x$ \\ \hline
$^{12}$C & $^{11}$B & $p$ & $1p_{3/2}$ & 1.688 \cite{Coh67}, 1.706 \cite{Rud01a} \\
$^{12}$C$^*_{4.44}$ & $^{11}$B & $p$ & $1p_{1/2}$ & $-0.741$\footnote{sign determined empirically} \cite{Coh67}, $-0.505$ \cite{Rud01a} \\
                    &          &     & $1p_{3/2}$ & $\pm 0.04$\footnote{Not used, sign not determined} \cite{Coh67}, $0.505$ \cite{Rud01a} \\
\hline
\end{tabular}
\caption{Spectroscopic amplitudes for the $\left<^{12}\mathrm{C} \mid \protect{^{11}\mathrm{B}} + p \right>$ overlaps used in the
CCBA calculations. Amplitudes $\mathrm{S}_x$ for the A = C + $x$ overlaps are given, where the $n\ell_j$ are the quantum numbers for
the relative motion of $x$ about the core C. Note that the signs of the amplitudes from Ref.\ \cite{Rud01a} are given here according
to the phase convention used in the {\sc fresco} code \cite{Tho88}. These differ from the signs given in Ref.\ \cite{Rud01a} by
the following factor: $(-1)^{J_C + j - J_A}$, where $J_C$ is the spin of the core, C, $J_A$ the spin of the composite, $C + x = A$, and
$j$ is the total relative angular momentum of $x$ with respect to $C$.}
\label{tab2}
\end{center}
\end{table}

Different choices of $^{11}$B + $^{209}$Bi optical potential parameters, e.g.\ those of Ref.\ \cite{Shr98}, did not
improve the agreement between the CCBA calculations and the stripping data.
The spectrum presented in Ref.\ \cite{Tot76} shows that transitions to levels of $^{209}$Bi leaving the
$^{11}$B in its 2.12-MeV $1/2^-$ excited state were weakly populated---almost on the level of background---and
inelastic couplings between the levels of $^{11}$B in the exit partition were therefore omitted from our calculations.

The calculations at the other energies used identical inputs with two exceptions: the entrance channel optical potential
parameters and the spectroscopic amplitudes for the $\left<^{209}\mathrm{Bi} \mid \protect{^{208}\mathrm{Pb}} + p \right>$ 
overlaps. The entrance channel optical potential parameters at 77.4 MeV were obtained by fitting the 76.5 MeV $^{12}$C +
$^{208}$Pb elastic scattering data of Rudakov {\em et al.\/} \cite{Rud01} and at 116.4 MeV by fitting the 118 MeV elastic
scattering data of Friedman {\em et al.\/} \cite{Fri72}. At 101 MeV the entrance channel potential was obtained by
fitting the elastic scattering angular distribution calculated using the optical model and the parameters of Ref.\
\cite{Tot76} with a coupled channel calculation. The resulting parameters are given in Table \ref{tab:omp}. The spectroscopic
amplitudes for the $\left<^{209}\mathrm{Bi} \mid \protect{^{208}\mathrm{Pb}} + p \right>$ overlaps were adjusted at each energy
to give the best description of the relevant stripping data. 

\subsection{The $^{208}$Pb($^{12}$C,$^{13}$C)$^{207}$Pb neutron pickup reaction}\label{calcs:n}

The data sets of Refs.\ \cite{Tot76} and \cite{Oer84} for this reaction were also reanalyzed by means of CCBA calculations
including the two-step transfer path via the 4.44-MeV $2^+$ excited state of $^{12}$C, and employing the prior form of the
DWBA formalism with the full complex remnant term. Inputs were similar to the corresponding FR-DWBA calculations of Ref.\
\cite{Tot76} and the entrance channel potentials and coupling strengths were identical to those described in the previous
subsection. The five-parameter $^{13}$C + $^{207}$Pb Woods-Saxon potential of Table I of Ref.\ \cite{Tot76} was used in the
exit channel. This was obtained by fitting $^{13}$C + $^{207}$Pb elastic scattering data at an appropriate energy, in this
case 86.1 MeV. The $^{12}\mathrm{C} + n$ and $^{207}\mathrm{Pb} + n$ binding potentials were as used in Ref.\ \cite{Tot76}. 

Two sets of spectroscopic amplitudes for the $\left<^{13}\mathrm{C} \mid \protect{^{12}\mathrm{C}(0^+)} + n \right>$ and 
$\left<^{13}\mathrm{C} \mid \protect{^{12}\mathrm{C}(2^+)} + n \right>$ overlaps were tested, those of Cohen and Kurath
\cite{Coh67} and those of Ziman {\it et al.\/} \cite{Zim97}. The values are given in Table \ref{tab3}.  
\begin{table}
\begin{center}
\begin{tabular}{|c|c|c|c|c|}
\hline
A & C & $x$ & $n\ell_j$ & $\mathrm{S}_x$ \\ \hline
$^{13}$C & $^{12}$C & $n$ & $1p_{1/2}$ & 0.783 \cite{Coh67}, 0.601 \cite{Zim97} \\
$^{13}$C & $^{12}$C$^*_{4.44}$ & $n$ & $1p_{3/2}$ & 1.059\footnote{sign determined empirically} \cite{Coh67}, 1.124 \cite{Zim97} \\
\hline
\end{tabular}
\caption{Spectroscopic amplitudes for the $\left<^{13}\mathrm{C} \mid \protect{^{12}\mathrm{C}} + n \right>$ overlaps used in the
CCBA calculations. Amplitudes $\mathrm{S}_x$ for the A = C + $x$ overlaps are given, where the $n\ell_j$ are the quantum numbers for
the relative motion of $x$ about the core C. Note that the signs of the amplitudes from Ref.\ \cite{Rud01a} are given here according
to the phase convention used in the {\sc fresco} code \cite{Tho88}. These differ from the signs given in Ref.\ \cite{Rud01a} by
the following factor: $(-1)^{J_C + j - J_A}$, where $J_C$ is the spin of the core, C, $J_A$ the spin of the composite, $C + x = A$, and
$j$ is the total relative angular momentum of $x$ with respect to $C$.}
\label{tab3}
\end{center}
\end{table}
The relative signs of the Cohen and Kurath \cite{Coh67} spectroscopic amplitudes for the $\left<^{13}\mathrm{C} \mid 
\protect{^{12}\mathrm{C}(0^+)} + n \right>$ and $\left<^{13}\mathrm{C} \mid \protect{^{12}\mathrm{C}(2^+)} + n \right>$ overlaps
were determined empirically, as described above for the corresponding spectroscopic amplitudes for the $^{208}$Pb($^{12}$C,$^{11}$B)$^{209}$Bi
proton stripping reaction. In this case the best description was obtained with a positive sign, a negative sign being found to destroy
the already good description of the neutron pickup data obtained with the FR-DWBA. Ziman {\it et al.\/} \cite{Zim97} give values
for the spectroscopic amplitudes which were used directly, after converting the signs to the phase convention used by {\sc fresco},
as described in the previous subsection. The spectroscopic amplitudes for the $\left<^{208}\mathrm{Pb} \mid \protect{^{207}\mathrm{Pb}} 
+ n \right>$ overlaps were adjusted at each incident $^{12}$C energy to give the best description of the relevant pickup data. 

\section{Results}\label{results}
\subsection{The $^{208}$Pb($^{12}$C,$^{11}$B)$^{209}$Bi proton stripping reaction}

\begin{figure}
\includegraphics[clip,width=10cm]{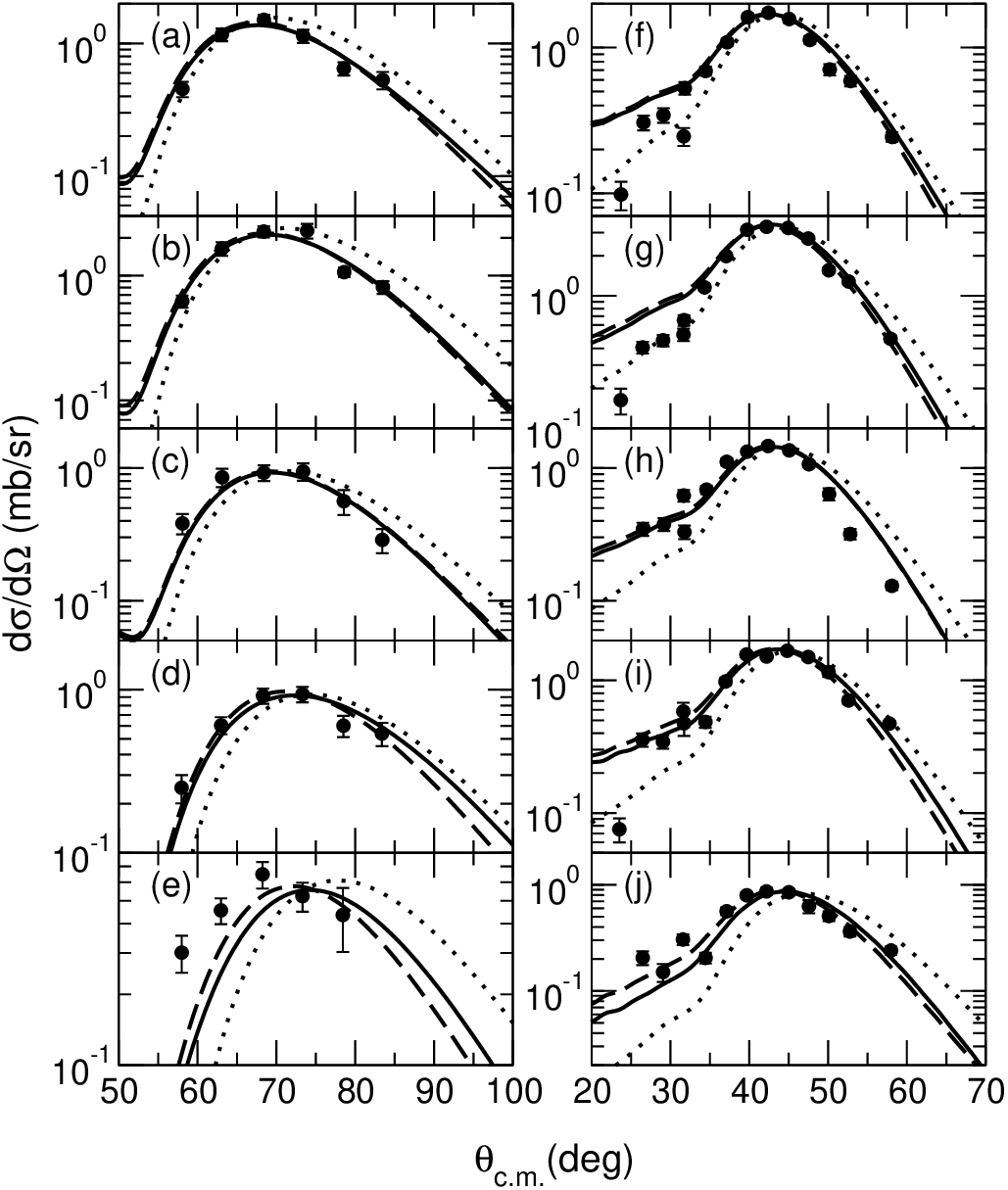}
\caption{Calculations for the $^{208}$Pb($^{12}$C,$^{11}$B)$^{209}$Bi reaction at bombarding energies of 77.4 MeV (left)
and 97.9 MeV (right) compared with the data of Ref.\ \cite{Tot76}; (a) to (e) denote stripping to the 0.0-MeV $9/2^-$,
0.90-MeV $7/2^-$, 1.61-MeV $13/2^+$, 2.84-MeV $5/2^-$, and 3.12-MeV $3/2^-$ states of $^{209}$Bi at a bombarding energy
of 77.4 MeV and (f) to (j) at a bombarding energy of 97.9 MeV. The solid curves denote the results of CCBA
calculations using projectile-overlap spectroscopic amplitudes derived from Cohen and Kurath \cite{Coh67}. The dotted
curves denote the results of DWBA calculations using the parameters of Ref.\ \cite{Tot76}. The dashed curves denote the
results of CCBA calculations using the projectile-overlap spectroscopic amplitudes of Ref.\ \cite{Rud01a}.}
\label{fig1}
\end{figure}

\begin{figure}
\includegraphics[clip,width=10cm]{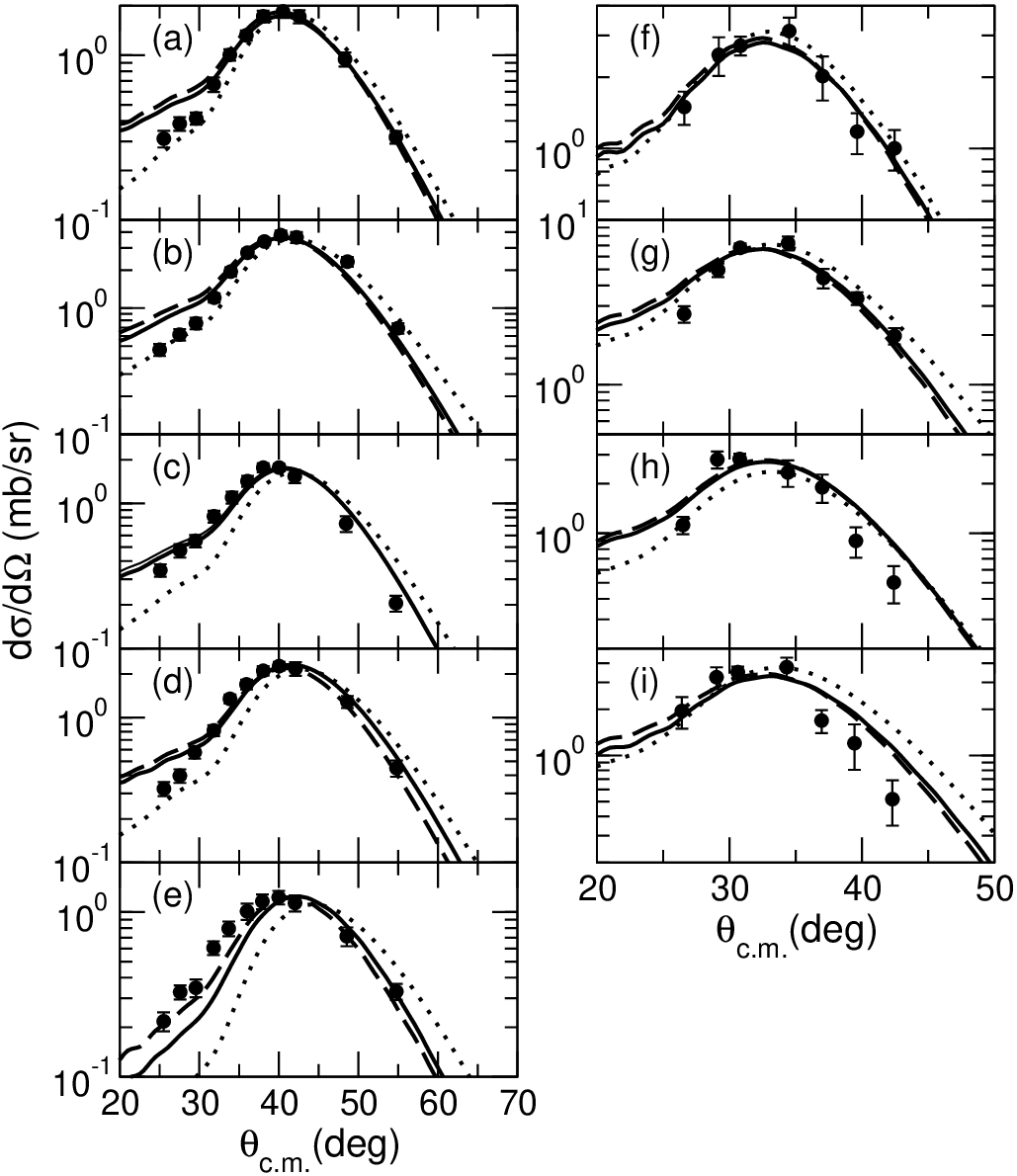}
\caption{Calculations for the $^{208}$Pb($^{12}$C,$^{11}$B)$^{209}$Bi reaction at bombarding energies of 101 MeV (left)
and 116.4 MeV (right) compared with the data of Refs.\ \cite{Oer84} and \cite{Tot76}, respectively; (a) to (e) denote stripping to the 0.0-MeV $9/2^-$,
0.90-MeV $7/2^-$, 1.61-MeV $13/2^+$, 2.84-MeV $5/2^-$, and 3.12-MeV $3/2^-$ states of $^{209}$Bi at a bombarding energy
of 101 MeV and (f) to (i) at a bombarding energy of 116.4 MeV. The solid curves denote the results of CCBA
calculations using projectile-overlap spectroscopic amplitudes derived from Cohen and Kurath \cite{Coh67}. The dotted
curves denote the results of DWBA calculations using the parameters of Ref.\ \cite{Tot76}. The dashed curves denote the
results of CCBA calculations using the projectile-overlap spectroscopic amplitudes of Ref.\ \cite{Rud01a}.}
\label{fig2}
\end{figure}

In Figs.\ \ref{fig1} and \ref{fig2} we compare the results of CCBA calculations employing the coupling scheme described in section
\ref{calcs:p} with the data for the $^{208}$Pb($^{12}$C,$^{11}$B)$^{209}$Bi reaction at 77.4 MeV \cite{Tot76}, 97.9 MeV \cite{Tot76},
101 MeV \cite{Oer84}, and 116.4 MeV \cite{Tot76}. The calculations using the $^{12}\mathrm{C} \rightarrow \protect{^{11}\mathrm{B}} + p$
spectroscopic amplitudes of Cohen and Kurath \cite{Coh67} are denoted by the solid curves and
those using the corresponding spectroscopic amplitudes of Ref.\ \cite{Rud01a} by the dashed curves. For most levels the results 
are virtually indistinguishable, exceptions being stripping to the 3.12-MeV $3/2^-$ and, to a lesser extent, the 2.84-MeV $5/2^-$
state, where the spectroscopic amplitudes of Ref.\ \cite{Rud01a} give an improved description of the data. While the goal of this
work is to investigate the proton stripping reaction mechanism rather than extract proton spectroscopic factors for the
$\left<^{209}\mathrm{Bi} \mid \protect{^{208}\mathrm{Pb}} + p \right>$ overlaps  we note that the values obtained using the
projectile-overlap spectroscopic amplitudes of Ref.\ \cite{Rud01a} show only minor variations from those obtained with the
Cohen and Kurath amplitudes.
Also shown are the results of DWBA calculations using the parameters of Ref.\ \cite{Tot76}. The
agreement of the CCBA calculations with the data is much better than that of the DWBA at all energies and for all
levels of $^{209}$Bi. Depending on the bombarding energy and the particular level of the $^{209}$Bi residual involved,
the inclusion of the two-step transfer path in the CCBA calculations improves the shape of the angular distribution 
and/or shifts the peak to smaller angles. The overall agreement of the CCBA calculations with the data is good, with the
exception of that for population of the 3.12-MeV $3/2^-$ state at all bombarding energies, the angular shift of the
calculated peaks being not quite enough to match the data for stripping to this state. 

A comparison of the spectroscopic amplitudes for the various $\left<^{12}\mathrm{C} \mid \protect{^{11}\mathrm{B}} + p \right>$ 
overlaps in Table \ref{tab2} reveals that while the values for the overlap linking the ground states of the core and composite nuclei
do not differ significantly between the two calculations, those linking the 4.44-MeV $2^+$ excited state of
the $^{12}$C composite with the ground state of the $^{11}$B core exhibit a substantial difference in the $1p_{3/2}$
components; that of Cohen and Kurath \cite{Coh67} being negligible whereas that of Ref.\ \cite{Rud01a}
is the same magnitude (but opposite sign) as that of the $1p_{1/2}$ component. Nevertheless, both sets give equivalent
descriptions of the ($^{12}$C,$^{11}$B) proton stripping data, with a slight preference for the  values of Ref.\
\cite{Rud01a}. Therefore, although it is clear that the two-step transfer path via the 4.44-MeV $2^+$
state of $^{12}$C is essential to a good description of the proton stripping data the reaction mechanism is not
sensitive to the precise nature of the associated projectile-like overlap.  

\subsection{The $^{208}$Pb($^{12}$C,$^{13}$C)$^{207}$Pb neutron pickup reaction}

\begin{figure}
\includegraphics[clip,width=10cm]{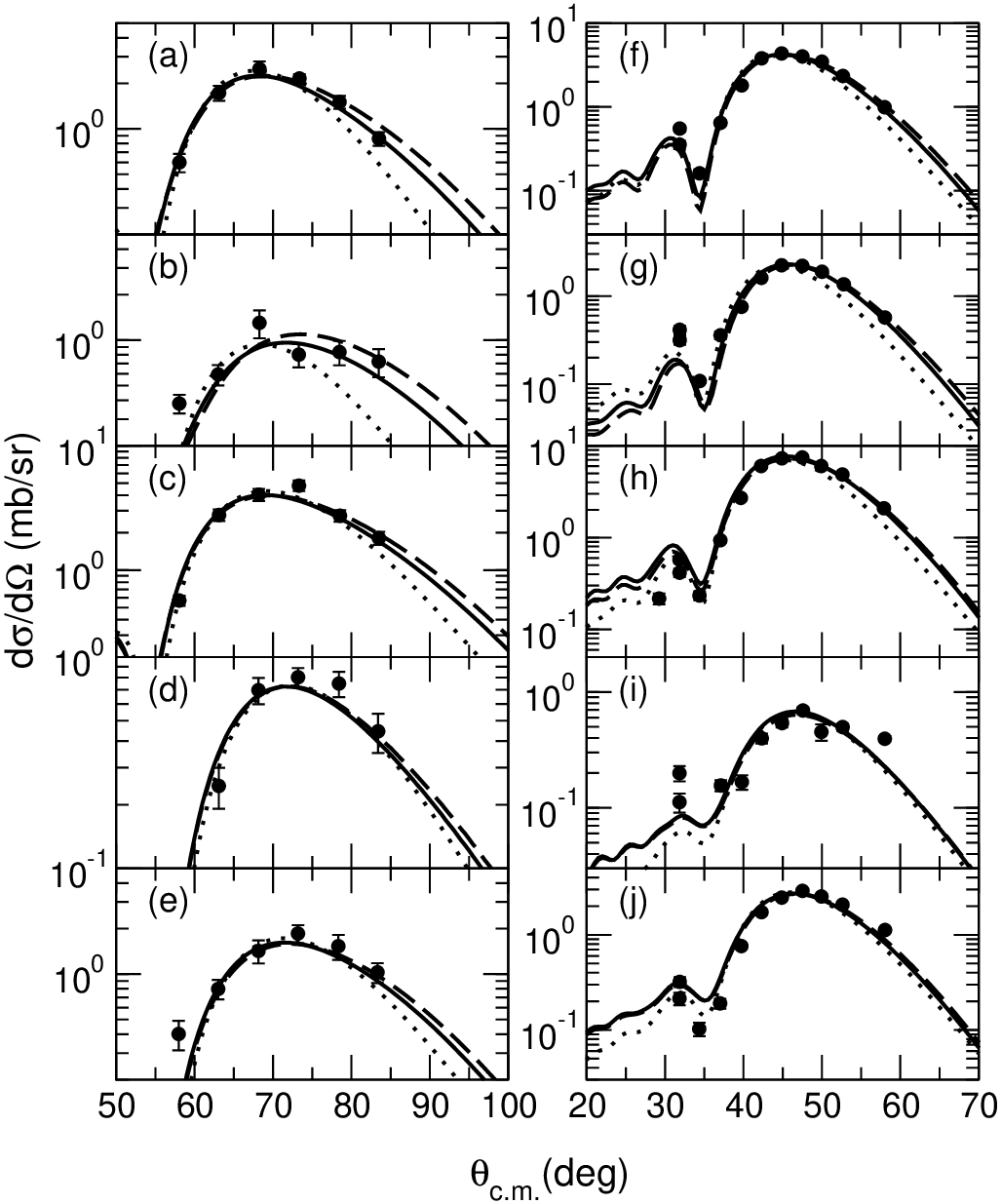}
\caption{Calculations for the $^{208}$Pb($^{12}$C,$^{13}$C)$^{207}$Pb reaction at bombarding energies of 77.4 MeV (left)
and 97.9 MeV (right) compared with the data of Ref.\ \cite{Tot76}; (a) to (e) denote pickup to the 0.0-MeV $1/2^-$,
0.57-MeV $5/2^-$, 0.90-MeV $3/2^-$, 1.63-MeV $13/2^+$, and 2.34-MeV $7/2^-$ states of $^{207}$Pb at a bombarding energy
of 77.4 MeV and (f) to (j) at a bombarding energy of 97.9 MeV. The solid curves denote the results of CCBA
calculations using projectile-overlap spectroscopic amplitudes derived from Cohen and Kurath \cite{Coh67}. The dotted
curves denote the results of DWBA calculations using the parameters of Ref.\ \cite{Tot76}. The dashed curves denote the
results of CCBA calculations using the projectile-overlap spectroscopic amplitudes of Ref.\ \cite{Zim97}.}
\label{fig3}
\end{figure}

\begin{figure}
\includegraphics[clip,width=10cm]{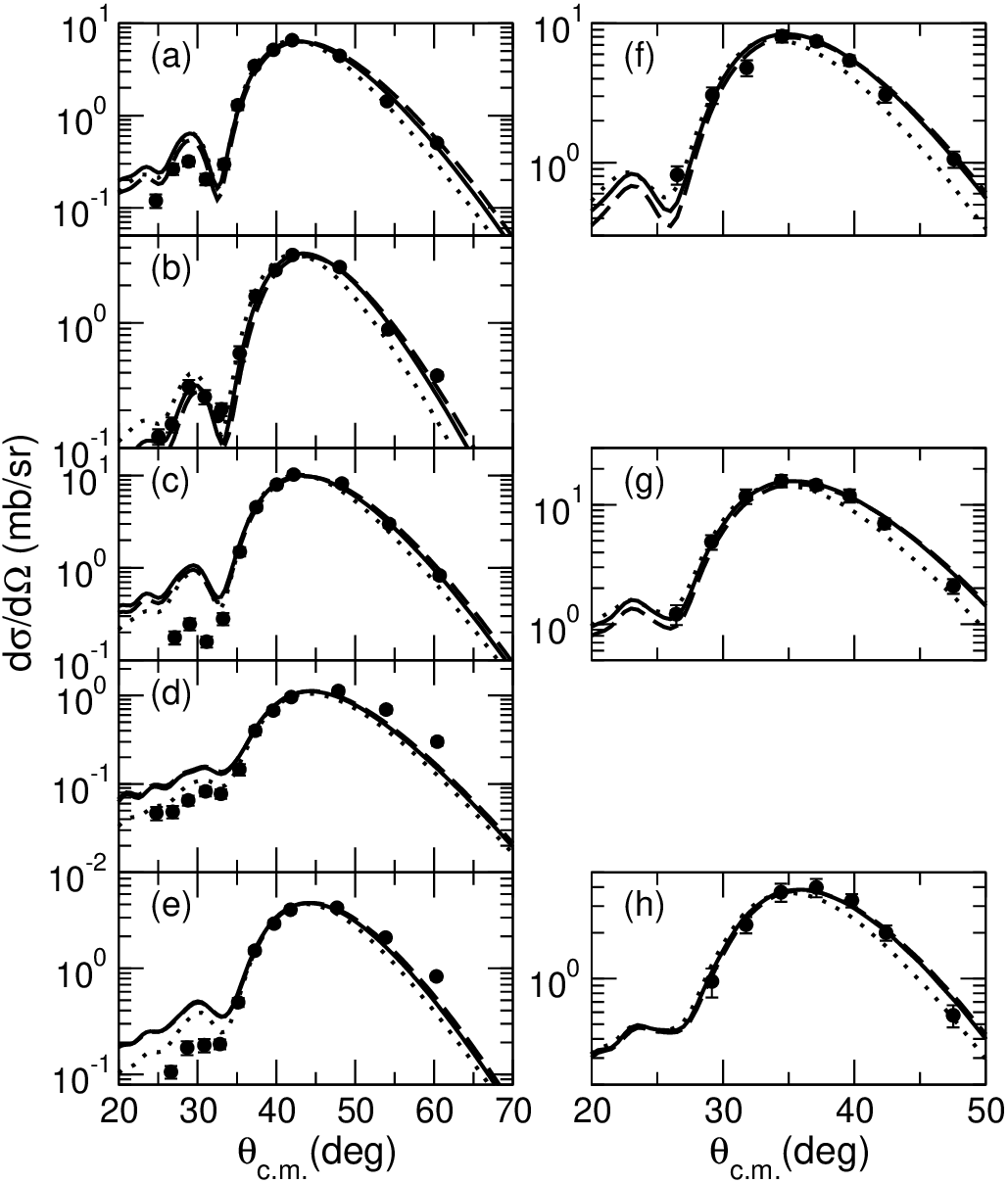}
\caption{Calculations for the $^{208}$Pb($^{12}$C,$^{13}$C)$^{207}$Pb reaction at bombarding energies of 101 MeV (left)
and 116.4 MeV (right) compared with the data of Refs.\ \cite{Oer84} and \cite{Tot76}, respectively; (a) to (e) denote pickup to the 0.0-MeV $1/2^-$,
0.57-MeV $5/2^-$, 0.90-MeV $3/2^-$, 1.63-MeV $13/2^+$, and 2.34-MeV $7/2^-$ states of $^{207}$Pb at a bombarding energy
of 101 MeV and (f) to (h) the 0.0-MeV $1/2^-$, 0.90-MeV $3/2^-$, and 2.34-MeV $7/2^-$ states of $^{207}$Pb at a 
bombarding energy of 116.4 MeV. The solid curves denote the results of CCBA
calculations using projectile-overlap spectroscopic amplitudes derived from Cohen and Kurath \cite{Coh67}. The dotted
curves denote the results of DWBA calculations using the parameters of Ref.\ \cite{Tot76}. The dashed curves denote the
results of CCBA calculations using the projectile-overlap spectroscopic amplitudes of Ref.\ \cite{Zim97}.}
\label{fig4}
\end{figure}

The results of CCBA calculations employing the coupling scheme described in section \ref{calcs:n} are compared with
the data for the $^{208}$Pb($^{12}$C,$^{13}$C)$^{207}$Pb neutron pickup reaction at 77.4 MeV \cite{Tot76}, 97.9 MeV
\cite{Tot76}, 101 MeV \cite{Oer84}, and 116.4 MeV \cite{Tot76} in Figs.\ \ref{fig3} and \ref{fig4}. The calculations
employing the $^{13}\mathrm{C} \rightarrow \protect{^{12}\mathrm{C}} + n$ spectroscopic amplitudes of Cohen and
Kurath \cite{Coh67} are denoted by the solid curves and those employing the values of Ziman {\it et al.\/} \cite{Zim97}
by the dashed curves.
Also shown are the results of FR-DWBA calculations using the parameters of Ref.\ \cite{Tot76}, denoted by the dotted curves.
The inclusion of the two-step transfer path via the $^{12}$C 4.44-MeV $2^+$ state in all cases improves the already good
description of the data by the FR-DWBA calculations, broadening slightly the peaks of the angular distributions, although
the effect is by no means as important as for the proton stripping reaction. Both sets
of $^{13}\mathrm{C} \rightarrow \protect{^{12}\mathrm{C}} + n$ spectroscopic amplitudes give similar results, although
use of the Ziman {\it et al.\/} \cite{Zim97} values consistently leads to larger spectroscopic amplitudes for the 
$\left<^{208}\mathrm{Pb} \mid \protect{^{207}\mathrm{Pb}} + n\right>$ overlaps, unlike for the $^{208}$Pb($^{12}$C,$^{11}$B)$^{209}$Bi
reaction where both sets of $^{12}\mathrm{C} \rightarrow \protect{^{11}\mathrm{B}} + p$ spectroscopic amplitudes yielded similar 
values for the $^{209}\mathrm{Bi} \rightarrow \protect{^{208}\mathrm{Pb}} + p$ amplitudes.
This merely reflects the fact that for the $^{208}$Pb($^{12}$C,$^{11}$B)$^{209}$Bi reaction the spectroscopic factors
(the squares of the spectroscopic amplitudes given in the tables) 
for the $\left<^{12}\mathrm{C}(0^+) \mid \protect{^{11}\mathrm{B}} + p 
\right>$ overlap of Refs.\ \cite{Coh67} and \cite{Rud01a} are almost identical, see Table \ref{tab2}, whereas for the
$^{208}$Pb($^{12}$C,$^{13}$C)$^{207}$Pb reaction the spectroscopic factor for the $\left<^{13}\mathrm{C} \mid 
\protect{^{12}\mathrm{C}(0^+)} + n\right>$ overlap of Cohen and Kurath \cite{Coh67} is approximately a factor of 1.7 greater
than that of Ziman {\it et al.} \cite{Zim97}, see Table \ref{tab3}. Both sets of spectroscopic amplitudes give equivalent
descriptions of the data so that it is not possible to choose between them, although the Cohen and Kurath amplitudes yield
values for the $^{208}\mathrm{Pb} \rightarrow \protect{^{207}\mathrm{Pb}} + n$ spectroscopic factors that are more
consistent with other determinations using light ion reactions, for example. 

\section{Summary and Conclusions}
In common with many heavy-ion induced proton transfer reactions the data for the $^{208}$Pb($^{12}$C,$^{11}$B)$^{209}$Bi
proton stripping reaction could not be satisfactorily described with FR-DWBA calculations without {\em ad hoc}
adjustments to the exit channel optical potentials \cite{Tot76,Oer84}, the calculations tending to peak at larger
angles than the data. It was remarked \cite{Tot76} that the discrepancy increased as the $Q$ value became more
negative (i.e.\ higher lying states in $^{209}$Bi were populated) or the bombarding energy was reduced, that is, as
the mismatch between the grazing angular momenta in the entrance and exit channels increased. It was speculated
\cite{Tot76,Oer84} that the shifts in the angular distributions might be due to coupling to inelastic channels,
specifically those of the projectile and/or ejectile, but to date this has never been demonstrated.   

In this work we have shown that CCBA calculations including two-step transfer via the 4.44-MeV $2^+$ excited state
of $^{12}$C provide a good description of the available data sets. Couplings to the ground state reorientation and
excitation of the 2.12-MeV $1/2^-$ excited state of $^{11}$B had a relatively minor influence. The influence of
the two-step path becomes more important as the excitation energy of the residual $^{209}$Bi increases and/or
the bombarding energy decreases, suggesting that this coupling is indeed compensating for the increased
angular momentum mismatch. Two sets of shell
model spectroscopic amplitudes for the $\left< ^{12}\mathrm{C} \mid \protect{^{11}\mathrm{B}} + p \right>$ overlaps 
were employed, one derived from Cohen and Kurath \cite{Coh67} and the other taken from the work of Rudchik {\em et al.\/}
\cite{Rud01a}, and both gave equivalent descriptions of the stripping data, with a slight preference for the
values of Rudchik {\em et al.}. While the amplitudes for the overlap linking the ground state of the $^{12}$C
composite with the ground state of the $^{11}$B core were not significantly different, this was not the case for those
linking the excited state of the composite with the ground state of the core, which have completely different
weightings for the $1p_{3/2}$ component in the two calculations. We thus conclude that while inclusion of the two-step
transfer path via the $^{12}$C 4.44-MeV $2^+$ state is essential for a good description of the data the calculations
are not sensitive to the details of the accompanying projectile-like overlap.  

It was also demonstrated that similar CCBA calculations for the $^{208}$Pb($^{12}$C,$^{13}$C)$^{207}$Pb reaction
were able to improve slightly the already good description of the data by the FR-DWBA \cite{Tot76,Oer84}, although
the influence of the two-step transfer via the 4.44-MeV $2^+$ excited state was significantly smaller than for
the proton-stripping case. Two sets of shell model spectroscopic amplitudes for the $\left< ^{13}\mathrm{C} \mid 
\protect{^{12}\mathrm{C}} + n \right>$ overlaps \cite{Coh67,Zim97} were tested, both giving equivalent descriptions
of the pickup data, although the values of Ziman {\it et al.\/} \cite{Zim97} yielded somewhat larger spectroscopic
amplitudes for the $\left<^{208}\mathrm{Pb} \mid \protect{^{207}\mathrm{Pb}} + n \right>$ overlaps. This is 
commensurate with the ratio of the spectroscopic {\it factors} (the squares of the spectroscopic amplitudes) for the
$\left< ^{13}\mathrm{C} \mid \protect{^{12}\mathrm{C}(0^+)} + n \right>$ overlap for the two sets, see Table \ref{tab3}. 
We again conclude that the calculations are not sensitive to the details of the projectile-like overlap, with the
exception that the relative positive sign of the amplitudes for the $\left<^{13}\mathrm{C} \mid \protect{^{12}\mathrm{C}(0^+)}
+ n\right>$ and $\left<^{13}\mathrm{C} \mid \protect{^{12}\mathrm{C}(2^+)} + n\right>$ overlaps is firmly established; a
relative negative sign shifts the transfer peaks to smaller angles by 1--2$^\circ$, destroying the agreement with the
data. 

\begin{figure}
\includegraphics[clip,width=10cm]{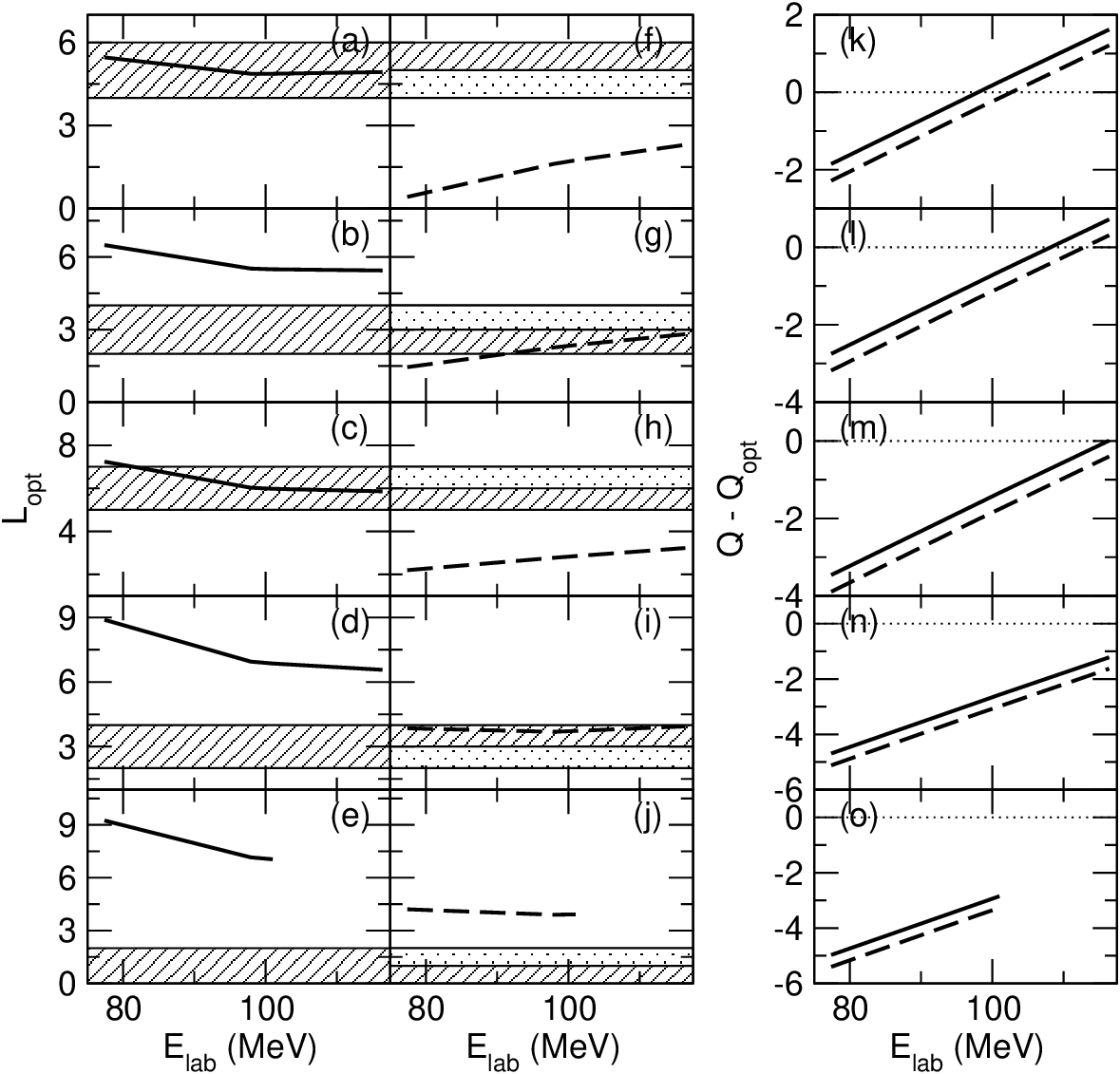}
\caption{For the $^{208}$Pb($^{12}$C,$^{11}$B)$^{209}$Bi reaction: In (a) to (e) the solid curves plot the optimum
angular momentum transfer ($L_\mathrm{opt}$) values as a function of $^{12}$C bombarding energy for stripping from
the $^{12}$C ground state and population of the 0.0-MeV $9/2^-$, 0.90-MeV $7/2^-$, 1.61-MeV $13/2^+$, 2.84-MeV $5/2^-$,
and 3.12-MeV $3/2^-$ states of $^{209}$Bi, respectively. The hatched bands denote the range of allowed $L$ transfers
for each transition. In (f) to (j) the dashed curves plot the $L_\mathrm{opt}$ values as a function of $^{12}$C bombarding 
energy for stripping from the $^{12}$C 4.44-MeV $2^+$ excited state. The hatched bands denote the range of allowed $L$ transfers
when the stripped proton is in the $1p_{3/2}$ level (as for stripping from the $^{12}$C ground state) and the dotted bands
when the stripped proton is in the $1p_{1/2}$ level. In (k) to (o) the solid curves plot $\mathrm{Q} - \mathrm{Q_{opt}}$
for stripping from the ground state of $^{12}$C and population of the 0.0-MeV $9/2^-$, 0.90-MeV $7/2^-$, 1.61-MeV $13/2^+$,
2.84-MeV $5/2^-$, and 3.12-MeV $3/2^-$ states of $^{209}$Bi, respectively while the dashed curves plot $\mathrm{Q} - \mathrm{Q_{opt}}$
for stripping from the $^{12}$C 4.44-MeV $2^+$ excited state.} 
\label{fig5}
\end{figure}

\begin{figure}
\includegraphics[clip,width=10cm]{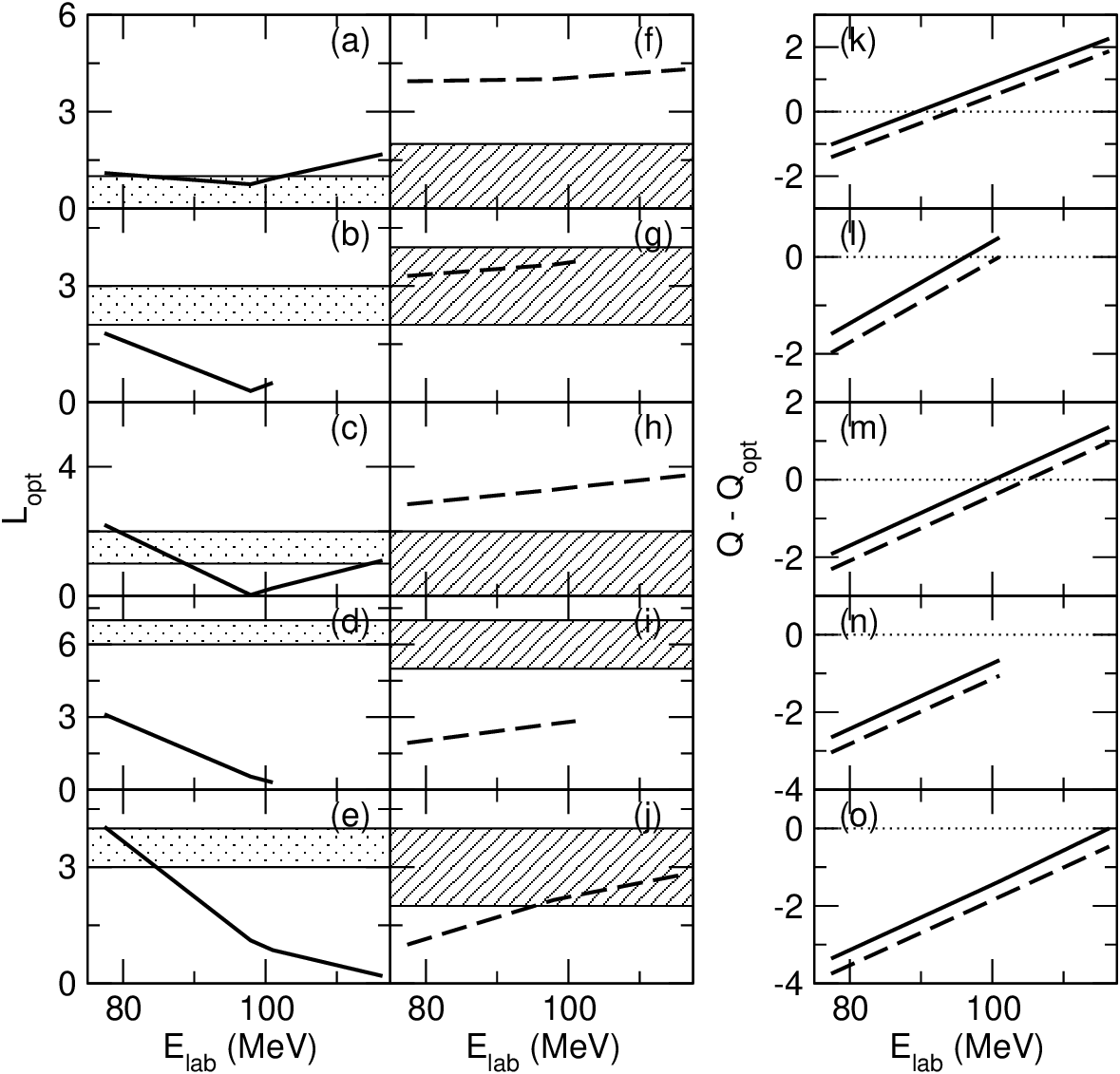}
\caption{For the $^{208}$Pb($^{12}$C,$^{13}$C)$^{207}$Pb reaction: In (a) to (e) the solid curves plot the optimum
angular momentum transfer ($L_\mathrm{opt}$) values as a function of $^{12}$C bombarding energy for pickup to
the $^{12}$C ground state and population of the 0.0-MeV $1/2^-$, 0.57-MeV $5/2^-$, 0.90-MeV $3/2^-$, 1.63-MeV $13/2^+$,
and 2.34-MeV $7/2^-$ states of $^{207}$Pb, respectively. The dotted bands denote the range of allowed $L$ transfers
for each transition. In (f) to (j) the dashed curves plot the $L_\mathrm{opt}$ values as a function of $^{12}$C bombarding
energy for pickup to the $^{12}$C 4.44-MeV $2^+$ excited state. The hatched bands denote the range of allowed $L$ transfers
for each transition. In (k) to (o) the solid curves plot $\mathrm{Q} - \mathrm{Q_{opt}}$
for pickup to the ground state of $^{12}$C and population of the 0.0-MeV $1/2^-$, 0.57-MeV $5/2^-$, 0.90-MeV $3/2^-$,
1.63-MeV $13/2^+$, and 2.34-MeV $7/2^-$ states of $^{207}$Pb, respectively while the dashed curves plot $\mathrm{Q} - \mathrm{Q_{opt}}$
for pickup to the $^{12}$C 4.44-MeV $2^+$ excited state.}
\label{fig6}
\end{figure}

In Figs.\ \ref{fig5} and \ref{fig6} we compare the optimum angular momentum transfer, $L_\mathrm{opt}$, with the allowed
angular momentum transfers and plot the values of $\mathrm{Q} - \mathrm{Q_{opt}}$ for each of the states populated in
the $^{208}$Pb($^{12}$C,$^{11}$B)$^{209}$Bi and $^{208}$Pb($^{12}$C,$^{13}$C)$^{207}$Pb reactions, respectively.  
Values of $\mathrm{Q_{opt}}$ for the $A(a,b)B$ transfer processes were calculated according to the Brink matching rules \cite{Bri72,For74}:
\begin{equation}
\label{eq:1}
\mathrm{Q_{opt}} = \left(Z_b Z_B - Z_a Z_A\right) e^2/R - \frac{1}{2} m v^2,
\label{eq:qopt}
\end{equation} 
where the charge on nucleus $i$ is denoted by $Z_i e$, the relative velocity of the two nuclei in the region of interaction 
(separated by distance $R$) by $v$, and the mass of the transferred particle by $m$. 
The relative velocity $v$ may be calculated as \cite{Olm78}:
\begin{equation}
v = \left[2\left(E_\mathrm{c.m.} - E_\mathrm{B}\right)/\mu\right]^{1/2},
\label{eq:v}
\end{equation}
where $E_\mathrm{B}$ and $\mu$ are the Coulomb barrier and reduced mass of the projectile-target system, respectively.
The optimum angular momentum transfers were calculated as the difference between the grazing angular momenta of the 
incoming ($L_a$) and outgoing ($L_b$) trajectories \cite{For74}:
\begin{equation}
L_\mathrm{opt} =\mid L_a - L_b \mid.
\end{equation}
The grazing angular momenta were taken as those values for which the elastic scattering $S$ matrix has $\protect{\mid S(L_a) \mid} \approx 
\protect{\mid S(L_b) \mid} \approx \protect{1/\sqrt{2}}$, obtained by smooth interpolation of the $S$ matrices calculated using the
appropriate optical potentials used in the FR-DWBA calculations. Following Ref.\ \cite{For74} the value of $R$ in eqn.\ (\ref{eq:qopt}) was  
taken as the strong absorption radius, defined as the distance of closest approach for a Rutherford trajectory with angular momentum
$L = L_a$. For the Coulomb barrier $E_\mathrm{B}$ in eqn.\ (\ref{eq:v}) we took a value of 57.4 MeV, the barrier height calculated using the
real part of the entrance channel optical potential used in the FR-DWBA calculations. 

Figures \ref{fig5} and \ref{fig6} show that while the $^{208}$Pb($^{12}$C,$^{13}$C)$^{207}$Pb neutron pickup is somewhat better
Q-matched than the $^{208}$Pb($^{12}$C,$^{11}$B)$^{209}$Bi proton stripping all the transfers considered fall within the likely
Q windows, with the possible exception of proton stripping populating the 2.84-MeV $5/2^-$ and 3.12-MeV $3/2^-$ states of $^{209}$Bi
at an incident $^{12}$C energy of 77.4 MeV, see Fig.\ \ref{fig5} (n) and (o), respectively. This also applies to transfers via the 
4.44-MeV $2^+$ excited state of $^{12}$C (plotted as the dashed curves) where $R$ and $v$ in eqn.\ (\ref{eq:qopt}) were calculated
using the appropriate lower value of $E_\mathrm{c.m.}$. In order to satisfy the angular momentum matching conditions the value
of $L_\mathrm{opt}$ for transfers leading to a particular final state of the residual nucleus should correspond closely to the
allowed $L$ transfers for the transition in question. These latter depend on both $\ell$ and $j$ of the transferred particle
with respect to the projectile-like and target-like cores. For proton stripping from the $^{12}$C ground state the proton is
in a pure $1p_{3/2}$ level, while for stripping from the 4.44-MeV $2^+$ excited state it is either in a pure $1p_{3/2}$ level
(Cohen and Kurath \cite{Coh67}; we ignored the small $1p_{1/2}$ level component of their wave function in our CCBA calculations) or a mixture of
$1p_{3/2}$ and $1p_{1/2}$ levels (Rudchik {\it et al.\/} \cite{Rud01}). For neutron pickup the transferred neutron is in a
$1p_{1/2}$ level ($^{12}$C in its ground state) or a $1p_{3/2}$ level ($^{12}$C in its 4.44-MeV $2^+$ excited state).  
The resulting ranges of allowed $L$ transfers have been plotted on Figs.\ \ref{fig5} and \ref{fig6} as the hatched ($1p_{3/2}$) and
dotted ($1p_{1/2}$) bands. While these accurately reflect the range of allowed values it should be recalled that these are in fact
quantized, not continuous. 

Figure \ref{fig5} only partially confirms the interpretation of Ref.\ \cite{Tot76} of the poor description of
the proton stripping data by the FR-DWBA as being due to poor angular momentum matching, since stripping from the $^{12}$C ground state
is seen to be well matched at all incident $^{12}$C energies for transitions leading to the 0.0-MeV $9/2^-$ and 1.61-MeV $13/2^+$ states
of $^{209}$Bi. However, for transitions to the remaining three levels which are poorly $L$ matched for stripping from the ground
state of $^{12}$C it is seen that stripping from the 4.44-MeV $2^+$ excited state is either well matched or better matched, at least
partially explaining the much improved description of the proton stripping data by the CCBA calculations. The situation is rather
less clear cut for the neutron pickup, see Fig.\ \ref{fig6}. Pickup for $^{12}$C in its ground state is reasonably well
$L$ matched for all transitions except that to the 1.63-MeV $13/2^+$ state of $^{207}$Pb but the description of the neutron pickup
by the FR-DWBA is noticeably better than the proton stripping even for cases where the $L$ matching is better for the latter.  
Also, while for some transitions pickup for $^{12}$C in its 4.44-MeV $2^+$ excited state is better $L$ matched than for the
ground state there appears to be little correlation between this and the importance of the two-step transfer path. We thus
conclude that while the matching conditions provide a useful aid to an understanding of the reaction mechanisms for heavy ion transfer data 
they do not tell the full story. Nevertheless, the matching concept remains an important tool in helping to explain why the DWBA 
fails for certain reactions. 

Similar discrepancies between the results of FR-DWBA calculations and the data exist for 
other proton transfer reactions, e.g.\ the $^{208}$Pb($^7$Li,$^6$He)$^{209}$Pb reaction
at 52 MeV \cite{Zel79}. In this case the $^7$Li projectile has strong ground state reorientation and coupling to
its bound first excited state which may influence the transfer reaction in a similar manner to the $^{12}$C
coupling included in this work. However, the phenomenon is not restricted to proton transfers, since the $^{208}$Pb($^{11}$B,$^{10}$B)$^{209}$Pb
data of Ref.\ \cite{For74} show a similar angular displacement compared to FR-DWBA calculations. This system may
well be a case where inelastic couplings in the ejectile are important, since $^{10}$B is strongly deformed. 
In light of our conclusions concerning the spectroscopic amplitudes for the overlaps
involving the 4.44-MeV excited state of $^{12}$C such analyses should provide fruitful ground for collaboration
with structure theorists, since the apparent lack of sensitivity to the details of the overlap would seem to
require the use of theoretical spectroscopic amplitudes for firm conclusions to be drawn.   

In summary, we have demonstrated the importance of two-step transfer via the 4.44-MeV excited state of $^{12}$C
in describing the $^{208}$Pb($^{12}$C,$^{11}$B)$^{209}$Bi single proton stripping and confirmed its
relatively minor influence on the $^{208}$Pb($^{12}$C,$^{13}$C)$^{207}$Pb single neutron pickup reaction. While the
concepts of angular momentum and Q matching proved to be useful aids to understanding this difference it was
demonstrated that they do not provide a complete explanation, so that while they can point to cases where multi-step
reaction paths are important they will not necessarily be able to provide an {\it a priori} conclusion as to
the particular paths required. Many similar
cases exist where the reaction is angular momentum---or possibly also $Q$---mismatched and inelastic excitations
of the projectile and/or the ejectile may be important in obtaining a good description of the transfer data. 
Such cases may be successfully handled within the framework of CCBA or coupled reaction channels theory provided the necessary
spectroscopic amplitudes are available.

\end{document}